%% file: main.tex
\documentclass[fleqn,usenatbib]{mnras}
\usepackage{newtxtext,newtxmath}
\usepackage[T1]{fontenc}
\DeclareRobustCommand{\VAN}[3]{#2}
\let\VANthebibliography\thebibliography
\def\thebibliography{\DeclareRobustCommand{\VAN}[3]{##3}\VANthebibliography}

\usepackage{tikz}
\usepackage{pgfplots, pgfplotstable, pgf-pie, subfigure, etoolbox}
\usepgfplotslibrary{groupplots}
\pgfplotsset{compat=1.17}
\usepackage{chemformula}

\usepackage{multirow}
\title[]{Oxygen depletion in giant planets with different formation histories}
\author[Fonte S. et al.]{
Fonte S.$^{1}$\thanks{E-mail: sergio.fonte@inaf.it},
Turrini D.$^{1,2}$,
Pacetti E.$^{1,3}$,
Schisano E.$^{1}$,
Molinari S.$^{1}$,
Polychroni D.$^{4}$,
Politi R.$^{1}$,
Changeat Q.$^{5,6}$
\\
$^{1}$ INAF-Istituto di Astrofisica e Planetologia Spaziali, Via del Fosso del Cavaliere 100, I-00133, Rome, Italy\\
$^{2}$ INAF-Osservatorio Astrofisico di Torino, Via Osservatorio 20, I-10025, Pino Torinese (TO), Italy\\
$^{3}$ Dipartimento di Fisica, Sapienza Università di Roma, Piazzale Aldo Moro 2, I-00185, Rome, Italy\\
$^{4}$ INAF-Osservatorio Astronomico di Trieste, Via Giambattista Tiepolo, 11, I-34131 Trieste (TS), Italy\\
$^{5}$ European Space Agency (ESA), ESA Office, Space Telescope Science Institute (STScI), 3700 San Martin Drive, Baltimore, MD 21218, USA\\
$^{6}$ Department of Physics and Astronomy, University College London, Gower St., London WC1E6BT, UK \\
}
\date{Accepted 2023 January 19. Received 2022 December 23; in original form 2022 October 19}
\pubyear{2022}
\pgfplotsset{
    discard if not/.style 2 args={
        x filter/.code={
            \edef\tempa{\thisrow{#1}}
            \edef\tempb{#2}
            \ifx\tempa\tempb
            \else
                
            \fi
        }
    }
}

\begin{document}
\label{firstpage}
\pagerange{\pageref{firstpage}--\pageref{lastpage}}
\maketitle

\input{abstract}

\input{introduction} 
\input{model} 
\input{results} 
\input{discussion} 
\input{conclusions}
\input{acknowledgements}
\input{dataavailabilitystatement}
\bibliographystyle{mnras}
\bibliography{references}
\appendix 
\input{appendix}

\bsp
\label{lastpage}
\end{document}

%% file: abstract.tex
\begin{abstract}
The atmospheric C/O ratio of exoplanets is widely used to constrain their formation. To guarantee that the C/O ratio provides robust information, we need to accurately quantify the amount of C and O in exoplanetary atmospheres. In the case of O, water and carbon monoxide are generally studied as the two key carriers. However, oxygen is a very reactive element and does not bind with carbon; depending on the temperature, it also binds to refractory elements. Estimating the amount of oxygen bound to refractory elements is therefore critical for unbiased estimates of the C/O ratio. In this work, we investigate the oxygen deficit due to refractory elements and its effects on the atmospheric C/O ratio of giant exoplanets as a function of their metallicity and equilibrium temperature. We model the composition of planetary atmospheres assuming chemical equilibrium and using as input physically justified elemental mixtures arising from detailed planet formation simulations. Our results show how the interplay between the atmospheric temperature and non-solar abundances of oxygen and refractory elements can sequester large fractions of oxygen, introducing significant biases in evaluating the C/O ratio when this effect is not accounted for. We apply our results to the case of Jupiter in the Solar System and show how the currently estimated water abundance points to a true oxygen abundance that is four times the solar one.
\end{abstract}


%% file: introduction.tex
\section{Introduction} \label{sec:intro}
Oxygen and carbon are two of the most cosmically abundant elements and, together, account for about 80\% of the budget of planet-building material within the circumstellar discs surrounding young stars \citep[e.g.][]{Asplund_2009,lodders2010,Palme_2014, oberg2021}. Oxygen and carbon are partitioned between the gas and dust of circumstellar discs depending on their local thermodynamic conditions \citep[e.g.][]{Fegley_2010,Palme_2014,eistrup2016,oberg2021}. Hotter regions are characterised by a larger presence of carbon and oxygen within the gas, although varying fractions of these elements are linked to refractory materials already in the innermost 1-2 au of circumstellar discs \citep{lodders2003,lodders2010,Fegley_2010,jura2014,Palme_2014,bergin2015,doyle2019}. Conversely, colder regions see increasing abundances of the two elements trapped in solids as ice. 

Due to their different volatility, carbon and oxygen are not sequestered into solids from the disc gas at the same rate. Thus, the C/O abundance ratios of both gas and solids in circumstellar discs vary with the distance from the host star \citep[e.g.][]{oberg2011,eistrup2016,oberg2021}. As the disc composition is imprinted into planets during their formation process, the atmospheric C/O ratio of giant planets provides constraints on where they formed within their native discs \citep[][see also  \citealt{Madhusudhan_2016,oberg2021,turrini2022} and references therein for recent discussions]{oberg2011}. To probe the formation process of giant planets, therefore, it is critically important to quantify as accurately as possible the amounts of carbon and oxygen in their atmospheres.

Observations by the Hubble Space Telescope and the Spitzer Space Telescope are mainly sensitive to H$_2$O and CO. It is, therefore, a common practice in exoplanetary studies to estimate oxygen from the measured abundances of those two molecules \citep{Lee_2013,Kreidberg_2014,Line_2014,McDonald_2019,Changeaat_2020,Line_2021,Spake_2021,Yui_2021,Evans_2022,Changeat_2022,Edwards_2022,ERS_2022}. This is typically done either via chemical equilibrium assumptions or via direct measurements of the abundance of those two tracers.

Oxygen, however, is a very reactive element and, as in the case of circumstellar discs, it does not bind only with carbon but also to refractory elements \citep{burrows1999,Fegley_2010}. For example, in planetary atmospheres characterised by solar composition about 22\% of oxygen is expected to be bound to the rock-forming refractory elements Si, Fe, and Mg at temperatures lower than 1200 K \citep{burrows1999,Fegley_2010}. Studies recovering the oxygen abundance using H$_2$O and CO only, or those assuming chemical equilibrium models that do not include refractory elements, are therefore subject to biases. 

As a case in point, the role of refractory elements in sequestering oxygen has been recently invoked by \citet{cridland2019} to partially explain the unexpectedly high value of the C/O ratio inferred for two hot-Neptunes (GJ 436 b and HAT-P-26 b), when compared to predictions for a synthetic population of planets. However, we still lack an in-depth understanding of the connection between this process and the planet formation process, as well as its implications for giant planets and their atmospheres.

To estimate the potential bias in C/O estimates arising from neglecting molecules other than H$_2$O and CO, we investigate the fraction of oxygen sequestered in refractory elements (\textit{oxygen deficit} in the following). Our study uses realistic models of giant planet atmospheres governed by equilibrium chemistry and explores multiple equilibrium temperatures and initial elemental abundances resulting from different formation and migration histories. Specifically, we consider hot and warm Jupiters that start their formation between 5 and 130 au from the host star and accrete both gas and planetesimals while migrating to their final orbits based on the planet formation simulations and compositional modelling from \cite{Turrini_2021} and \cite{pacetti2022} (see Appendix \ref{appendixa} for details).

These planet formation scenarios result in planetary compositions enriched in refractory elements with respect to giant planets whose growth tracks are dominated by the accretion of nebular gas \citep{schneider2021,Turrini_2021,pacetti2022}.
We focus on the atmospheric layers with optical depths that are optimally accessible by the spectrometers on-boards the NASA/ESA/CSA James Webb Space Telescope (JWST, \citealt{Greene_2016}) and the ESA mission Ariel \citep{tinetti2018,tinetti2021}, but similar predictions can be easily produced for other observing conditions.

The rest of the paper is organised as follows. In section \ref{sec:model} we illustrate the thermo-physical and chemical modelling as well as the initial elemental compositions of the planetary atmospheres we investigate. In section \ref{sec:results} we show the partition of oxygen among its main carriers as a function of the planetary metallicity and equilibrium temperature. In section \ref{sec:discussion} we discuss the impact of the oxygen deficit for warm-to-hot Jupiters and for Jupiter in our own Solar system. We summarise our conclusions in section \ref{sec:conclutions}.

%% file: model.tex
\section{Model} \label{sec:model}

In this study, we model the composition of the exoplanetary atmospheres of warm and hot Jupiters under the assumption of chemical equilibrium. The atmospheres we model are characterised by different temperatures, metallicity values and elemental compositions. The input metallicity values and elemental compositions of the giant planets are obtained from the planet formation simulations of \cite{Turrini_2021}, hereafter Paper I. In the following we provide the key aspects of our model and its input data.

\subsection{Atmospheric modelling}
\label{atmo_model}

We use FastChem (\citealt{Stock_2018}) to solve the system of coupled nonlinear algebraic equations describing the atmospheric chemical equilibrium. The chemical network that FastChem implements is appropriate for temperatures in excess of 100~K, so it provides a reliable treatment in the range of atmospheric temperatures of warm and hot Jupiters (700--1500 K) we model in this work. As our focus is on quantifying the amount of oxygen sequestered by refractories, in this work we do not model the physical state of the resulting oxides, i.e. whether they remain in the gas phase or condense and trigger cloud formation. We defer the exploration of these aspects to future works.

As discussed by \cite{Stock_2018}, the temperature dependence of the dimensionless mass action constant for each chemical reaction $i$ in FastChem's chemical network is approximated as:

\begin{equation}\label{fig:kapprox}
\ln\overline{K}_i(T) = \frac{a_{0,i}}{T} + a_{1,i} \ln T + b_{0,i} + b_{1,i} T + b_{2,i} T^2
\end{equation}

with the coefficients $a_{k,i}$ and $b_{k,i}$ provided by FastChem in tabular form.

The thermo-physical state of the atmosphere is defined through its pressure-temperature relationship following the approach in \cite{Guillot_2010}. The temperature T is expressed as a function of the atmosphere's optical depth $\tau$ via 

\begin{equation}
            \overline{T^4} = \frac{3}{4}T^4_{int}\left( \frac{2}{3}+\tau\right) + \frac{3}{4}T^4_{eq}\left(\frac{2}{3} + a_1 + a_2 \right) 
\end{equation}

where $T_{int}$ is the temperature at the base of the atmosphere, which is assumed constant at 100 K \citep{Guillot_2010}, and $T_{eq}$ is the equilibrium temperature of the planet that depends on the planet-star distance $D$ and the star's temperature $T_{star}$ as discussed below. The quantities $a_1$ and $a_2$ incorporate the complex relationship between the atmosphere's optical depth and its opacity at thermal and visible wavelengths \citep{Guillot_2010}.

Following \cite{Guillot_2010}, we assume plane-parallel geometry and hydrostatic equilibrium to describe the atmosphere. Under these conditions, the local pressure $P$ and optical depth ${\tau}$ are linked by the following relation:

\begin{equation}
    P = \frac{g\tau}{k_{th}}
\end{equation}

where $g$ is gravity acceleration and $k_{th}$ is the opacity in the visible, for which we adopt a fiducial value of 0.01 cm$^{2}$/g again following \cite{Guillot_2010}.

\begin{table}
\caption{HAT-P-5b's characteristics \citep{Thorngren_2019}}
\label{tab:exoplanet}
\begin{tabular}{lll} \\ \hline
{Parameter}  & {Value} & {Unit}\\
\hline
$M_{b}$     & 0.98             & $M_J$       \\
$R_{b}$     & 1.21             & $R_J$       \\
\hline
$T_{\star}$          & 5960             & K           \\
$M_{\star}$          & 1.163            & $M_{\odot}$       \\
$R_{\star}$          & 1.137            & $R_{\odot}$       \\
\hline
\end{tabular}
\end{table}

We use HAT-P-5b and its host stars as templates on which to set the planetary and stellar parameters. HAT-P-5b's characteristics match well those expected for an older version of the newly formed, hot and expanded giant planet simulated in Paper I (1 M$_\textrm{J}$ and 1.6 R$_\textrm{J}$) after it undergoes secular cooling and shrinking. The main input parameters of HAT-P-5b and its star are derived from \cite{Thorngren_2019} and summarised in Tab.~\ref{tab:exoplanet}.

Since we are interested in exploring a range of equilibrium temperatures, we vary the orbital distance $D$ of the giant planet from the host star between 0.2 to 0.04 AU.  

This results in increasing planetary equilibrium temperatures $T_{eq}$ spanning from 700 to 1500 K as derived from

\begin{equation}\label{formula:teq}
    T_{eq} = T_{\star}\sqrt{\frac{R_{\star}}{2D}}.
\end{equation}

With these assumptions, we generate the set of eight different pressure-temperature profiles reported in Fig. \ref{fig:atmprofile} that we feed to FastChem for each of the six sets of elemental abundances in Tab.~\ref{tab:abscenarios}. We focus our analysis on the atmospheric layer encompassing the pressure range between 0.01 and 1 bar (see Fig. \ref{fig:atmprofile}) as it is the layer where both JWST \citep{Greene_2016} and the ESA mission Ariel \citep{tinetti2018,tinetti2021} have the optimal sensitivity.

\subsection{Elemental composition of the giant planets}
\label{planet_model}

To model the chemical initial conditions in the atmosphere, we consider six planetary mixtures resulting from the concurrent accretion of gas and planetesimals by a growing and migrating giant planet in the midplane of a protoplanetary disc. The disc compositional model assumes solar abundances \citep{Asplund_2009,Scott_I_2015,Scott_II_2015} and accounts for the presence of gas, ices, organics and refractories in the disc's midplane. The input elemental abundances in the planetary atmosphere are listed in Tab.~\ref{tab:abscenarios} and are the outcomes of the six planet formation simulations from Paper~I, coupled with the updated disc compositional model by \cite{pacetti2022}, hereafter Paper~II. We refer interested readers to Appendix~\ref{appendixa} for more details.

The six formation scenarios of Paper I simulate the growth and migration process of giant planets starting at different distances from the host star, namely between 5 and 130 au, and ending their formation at 0.4 au (see Tab.~\ref{tab:abscenarios} and Appendix \ref{appendixa}). The bulk metallicity of the giant planets increases with their initial planet formation distance (see Tab.~\ref{tab:elemental_enrichments}), as the giant planets migrate across larger fractions of the circumstellar disc and encounter more solid material to accrete. We assumed that the accreted gas and solids are split into the composing elements due to the high temperature of the young giant planet \citep{lissauer2009,dangelo2021} and recombine into molecules in its atmosphere. In the following, we will identify the six chemical mixtures based on their total bulk metallicity. The larger the migration, the more snowlines the giant planet crosses while migrating. As a result, the giant planets possess different abundances of C, O and refractory elements in the six formation scenarios.

The disc compositional model of Papers~I and II focuses on the four cosmically abundant elements nitrogen (N), carbon (C), oxygen (O), and sulphur (S), here reported in order of decreasing volatility. In these works, N, C, and O are partitioned in the disc midplane between refractory solids (rocks and metals), organics, ices, and gas. Their radial abundance profiles are based on the outcome of astrochemical models and on observational constraints provided by meteorites, comets, polluted white dwarfs, and the interstellar medium (see Appendix \ref{appendixa}, \citealt{oberg2021} and \citealt{turrini2022} for discussion). While N, C, and O are partitioned between the gas and solid phase across the disc, the available observational evidence suggests that the bulk of S is sequestered into refractory solids close to the host star (see Papers I and II and \citealt{Fegley_2010}, \citealt{kama2019} and \citealt{turrini2022} for discussion). Following Paper I and II, we adopt S as a proxy for all refractory elements and derive the planetary abundance of each refractory element X by multiplying the S abundance in the simulated giant planet by the stellar X/S abundance ratio. This approach allows us to account for the 25 most abundant heavy elements (see Tab.~\ref{tab:abscenarios}).

In Tab.~\ref{tab:elemental_enrichments} we report, for each of the six initial planet formation distances reported in Tab.~\ref{tab:abscenarios}, the total metallicity Z, and the abundances of C, O and refractory elements. In terms of refractory elements, we focus in particular on Fe, Mg and Si, as after C, O and N they provide the largest mass contribution to heavy elements \citep{lodders2010}. Both the metallicity and the elemental abundances in Tab.~\ref{tab:elemental_enrichments} are normalised to the relevant solar values. As can be immediately seen, refractory elements increase faster than O and C with increasing migration. 
The elemental compositions of the giant planets, therefore,  significantly deviate from the solar composition in terms of elemental abundance ratios (i.e. different elements show different enrichments), as discussed in Papers I and II.

\begin{table}
\caption{Elemental abundances of 25 elements in the atmosphere of the giant planet, resulting from the planet formation simulations from \citealt{Turrini_2021}, using the updated compositional model of the protoplanetary disc by \citealt{pacetti2022}. The elemental abundances are sorted by planetary migration scenario and expressed in dex (logarithmic abundance of atoms of a given element for every $10^{12}$ hydrogen atoms, see \citealt{Asplund_2009} and \citealt{lodders2010}).}
    \label{tab:abscenarios}
    \begin{tabular}{ccccccc}
    \hline
{Element} & \multicolumn{6}{c}{Initial distance of the planet (AU)} \\
 & {5} & {12} & {19} & {50} & {100} & {130} \\
\hline
 Al	& 6.38	 & 6.57	  & 7.15   & 7.06  & 7.21  & 7.42  \\
 Ar	& 6.40	 & 6.40	  & 6.40   & 6.40  & 6.40  & 6.40  \\
 C	& 8.44	 & 9.00	  & 9.12   & 9.39  & 9.14  & 9.35  \\
 Ca	& 6.27	 & 6.46	  & 7.04   & 7.35  & 7.10  & 7.31  \\
 Cl	& 6.15	 & 6.34	  & 6.52   & 7.23  & 7.38  & 7.19  \\
 Co	& 5.25	 & 5.03	  & 5.21   & 5.52  & 6.08  & 6.28  \\
 Cr	& 5.54	 & 6.12	  & 6.30   & 6.21  & 6.37  & 6.57  \\
 Cu	& 4.11	 & 4.29	  & 4.47   & 5.18  & 5.34  & 5.14  \\
 F	& 4.35	 & 4.54	  & 5.12   & 5.03  & 5.19  & 5.39  \\
 Fe	& 7.43	 & 8.01	  & 8.19   & 8.10  & 8.25  & 8.46  \\
 Ge	& 3.57	 & 4.15	  & 4.33   & 4.24  & 4.40  & 5.00  \\
 K	& 5.36	 & 5.14	  & 5.32  & 6.03  & 6.19  & 6.39  \\
 Mg	& 7.54	 & 8.13	  & 8.31  & 8.22  & 8.37  & 8.58  \\
 Mn	& 5.34	 & 5.52	  & 6.10  & 6.01  & 6.17  & 6.37  \\
 N	& 8.26	 & 8.30	  & 8.39  & 8.15  & 8.26  & 8.43  \\
 Na	& 6.16	 & 6.35	  & 6.53  & 7.24  & 7.39  & 7.20  \\
 Ni	& 6.12	 & 6.30	  & 6.48  & 7.19  & 7.35  & 7.15  \\
 O	& 9.15	 & 9.24	  & 9.00  & 9.29  & 9.45  & 10.05 \\
 P	& 5.36	 & 5.55	  & 6.13  & 6.04  & 6.19  & 6.40  \\
 S	& 7.07	 & 7.26	  & 7.44  & 8.15  & 8.30  & 8.11  \\
 Sc	& 3.08	 & 3.26	  & 3.44  & 4.15  & 4.31  & 4.11  \\
 Si	& 7.46	 & 8.05	  & 8.23  & 8.14  & 8.29  & 8.50  \\
 Ti	& 5.28	 & 5.07	  & 5.25  & 5.56  & 6.11  & 6.32  \\
 V	& 4.21	 & 4.39	  & 4.17  & 4.48  & 5.04  & 5.24  \\
 Zn	& 4.48	 & 5.06	  & 5.24  & 5.15  & 5.31  & 5.51  \\
 \hline
    \end{tabular}
\end{table}

\begin{table}
    \centering
    \caption{Total metallicity (Z) and enrichments in C, O, and refractory elements (among which Fe, Mg, and Si are the most abundant) of the atmospheres of the giant planets in the six formation scenarios simulated by \citealt{Turrini_2021}. The metallicity and the enrichments are expressed in units of the respective solar values. In this scale, a value of 1 indicates a perfect match with the corresponding solar quantity.}
    \label{tab:elemental_enrichments}
    \begin{tabular}{ccccc} \\
    \hline
    Initial Distance & $Z$ & C & O & Refractories \\
    (AU) & & & & (Fe/Mg/Si) \\
    \hline
    5 & 1.0 &   0.9 &   1.1 &	0.8 \\
    12 & 1.3 &   1.3 &   1.3	&   1.2 \\
    19 & 1.8 &   1.8 &   1.9	&   1.8 \\
    50 & 3.4 &   3.2	&   3.7	&   3.7 \\
    100 & 4.8 &   4.7	&   5.2	&   5.3 \\
    130 & 7.6 &   7.3	&   8.3	&   8.7 \\
    \hline
    \end{tabular}
\end{table}

\input{atmprofiles}

%% file: atmprofiles.tex
\begin{figure}
    \includegraphics[]{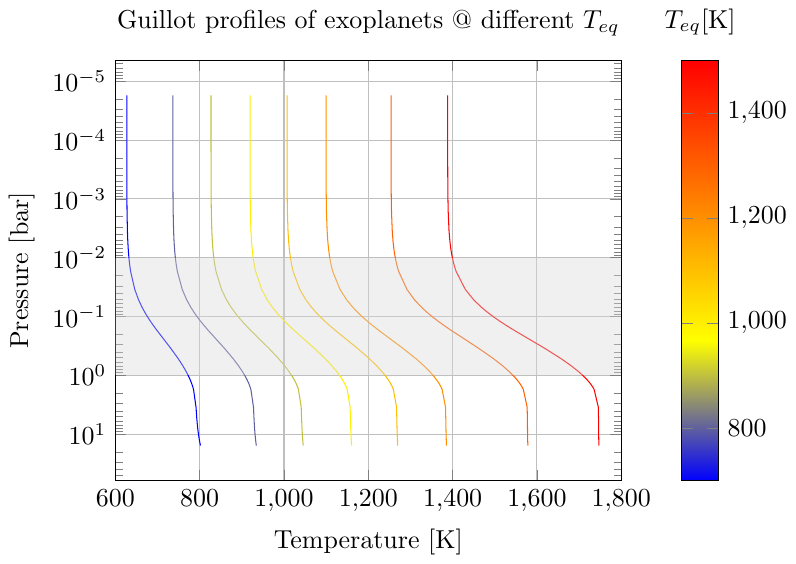}
    \caption{P-T profiles of the simulated giant planets for the eight orbital distances $D$ and equilibrium temperatures $T_{eq}$ we considered in this study. The grey region indicates the pressure range our atmospheric modelling focuses on, which has been chosen based on the atmospheric layer of highest sensitivity of the ESA mission Ariel \citep{tinetti2018,tinetti2021} and NASA/ESA/CSA JWST mission \citep{Greene_2016}}
    \label{fig:atmprofile}
\end{figure}

%% file: results.tex
\section{Results} \label{sec:results}


\input{exoplanetdef}

In this section, we discuss the abundances of all O-bearing molecules resulting from our atmospheric modelling with FastChem. The atmospheric models are computed considering eight planetary equilibrium temperatures spanning the range between 700 and 1500 K, and six formation and migration scenarios of the giant planet spanning initial formation distances between 5 
and 130 au. 
The resulting 48 atmospheric models are shown in Figs. \ref{fig:oxygendist1}, \ref{fig:oxygendist2} and \ref{fig:oxygendist3}. 

Each panel in these figures reports the molecular abundances resulting from FastChem for the specific equilibrium temperature as coloured bar charts. The different bar charts in each panel illustrate the distribution of O in the various planet formation scenarios. The planet formation scenarios are identified by their normalised metallicity value Z=Z$_{p}$/Z$_{*}$, where Z$_p$ and Z$_*$ are the planetary and stellar metallicity (see \citealt{thorngren2016} and Paper I) and Z goes from 1 to 7.6. Individual molecules are explicitly reported only if their contribution in sequestering O exceeds 1\% of total O; species not fulfilling this condition are grouped and their total contribution is reported under the label ``Other''.

In the following subsections, we will separately discuss the results for three classes of planetary equilibrium temperatures: \textit{warm} (700K$\leq T_{eq} \leq$800K), \textit{transitional hot} (900K$\leq T_{eq} \leq$1100K) and \textit{hot} (1200K$\leq T_{eq} \leq$1500K) planets, as the three categories show different properties in terms of chemical behaviour. As illustrated by Fig. \ref{fig:exoplanetdef}, the ``transitional hot'' label refers to the temperature range separating the two regimes where oxygen is carried by different carriers. Specifically, oxygen is locked mostly in water and refractories for "warm" planets, while it is carried preferentially by CO, water and SiO in the atmospheres of ``hot'' planets.

In the pressure region considered in the present study (0.01-1 bar, see Sect. \ref{sec:model}), the balance between the major C-bearing molecules CO and \ch{CH_4} is set by the reaction \ch{CO} + 3\ch{H2} $\rightleftharpoons$ \ch{CH4} + \ch{H2O} and favours CH$_4$ at the lowest temperatures we model ($\leq$800 K). For growing planetary temperatures the balance of the reaction gradually shifts in favour of CO production. Around 900 K the two molecules CO and \ch{CH4} equally contribute as C carriers in a gas with solar composition. At higher temperatures ($\geq$1000 K) CO is about 90\% of the C-bearing blend. We refer readers to \cite{Lodders_2002}, \cite{Fegley_2010} and \cite{Madhusudhan_2016} for more detailed discussions.

\subsection{Warm planets: 700K$\leq T_{eq} \leq$800K}\label{p_warm}

\input{oxygendist-700K-800K}

In the temperature regime of warm giant planets the main carriers of O are water and refractories (see Fig. \ref{fig:oxygendist1}). For increasing planetary metallicity values, the fraction of O incorporated into \ch{H2O} drops from the initial value of about 3/4 (73\%, see the cases Z=1 in Fig. \ref{fig:oxygendist1}) to less than 2/3 (62-63\%, see the cases Z=7.6 in Fig. \ref{fig:oxygendist1}) of total O. Most of this decrease occurs as soon as the metallicity Z shifts from stellar to super-stellar (i.e. between Z=1 and Z=1.3, see Fig. \ref{fig:oxygendist1})

This decrease in the role of water as a carrier of O is due to the faster increase of refractory elements with respect to O shown in Tab.~\ref{tab:elemental_enrichments}, as refractory elements (Fe-Mg-Si) increase by 50\% when going from Z=1 to 1.3 while O grows only by 18\% due to the different efficiencies with which gas and solids are accreted by the giant planet (see Paper I and II for detailed discussions). Among refractories, Fe sequesters between 10-13\% of total O, Mg between 12-17\%, while Si's contribution is mostly constant at 5-6\% of total O. 

When considering the physically-justified planetary compositions from Paper I, we find that 33-38\% of O is bound to refractory elements as soon as the planetary metallicity is super-stellar ($Z>1$). This value is significantly higher than the expected 22\% arising when solar abundance ratios are assumed between oxygen and refractories \citep{burrows1999,lodders2003,Fegley_2010}. In the case of Z=1, moreover, we find that refractories account for 27\% of the planetary oxygen as the different elements are not in solar proportions (see Tab.~\ref{tab:elemental_enrichments}). This highlights how the assumption of solar composition introduces biases in the interpretation of giant planet atmospheres.

\subsection{Transitional hot planets: 900K$\leq T_{eq} \leq$1100K}\label{p_trans}

\input{oxygendist-900K-1100K}

In this temperature range, planetary atmospheres exhibit a more complex behaviour than their colder counterparts discussed above. As shown in Fig. \ref{fig:oxygendist2}, the amount of oxygen sequestered by refractories is a function of both $T_{eq}$ and the metallicity Z (hence, the formation distance and migration of the growing giant planets). 

Moving toward hotter temperatures, transitional hot giant planets experience the expected shift from \ch{H2O} to \ch{CO} as the dominant carrier of O (see Fig. \ref{fig:oxygendist2}). In parallel, the fraction of O that is trapped by refractories undergoes a more radical change. At a fixed temperature $T_{eq}$, the amount of O linked to refractories increases with the planetary metallicity Z (see Fig. \ref{fig:oxygendist2}a, b, and c). For each metallicity Z, however, the role of refractories as carriers of O drastically decreases with increasing temperatures.

For $T_{eq}$=900 K, refractories sequester between 18\% and 36\% of total O when going from Z=1 to Z=7.6 largely due to the contributions of Fe and Mg (see Fig. \ref{fig:oxygendist2}a). Moving to $T_{eq}$=1000 K, the amount of oxygen trapped by refractories drops by a factor comprised between 3 for the lowest values of Z and 1.5-2 for the highest one. This decrease is due to the shrinking role of Fe and Mg (see Fig. \ref{fig:oxygendist2}b), while \ch{SiO} accounts for an almost constant fraction of 5-6\% of total O as in the case of warm giant planets. By $T_{eq}$=1100 K, refractories account for only 5-10\% of total O with Si becoming the main refractory O carrier (see Fig. \ref{fig:oxygendist2}c).

\subsection{Hot planets: 1200K$\leq T_{eq} \leq$1500K}\label{p_hot}

\input{oxygendist-1200K-1500K}

Hot giant planets show simpler behaviour than transitional hot ones. The volatile molecules \ch{CO} and \ch{H2O} play the leading role as O-bearing species, with \ch{CO} marginally dominant over \ch{H2O}. The role of refractories is limited to 5-8\% and is by large dominated by the constant 5-6\% contribution of \ch{SiO}, with only 0.5-2\% being cumulatively contributed by all remaining refractory elements. 

In the case of hot giant planets, estimating the atmospheric C/O ratio by measuring the abundance of O through \ch{CO} and \ch{H2O} proves more reliable. The measures are affected by a limited systematic error of the order of 6\% due to the neglected contribution of refractory elements. However, the interplay between the non-stellar composition of giant planets and the sequestration of O by refractories means that the partition of O between \ch{CO} and \ch{H2O} deviates from the expected picture even at such high temperatures, as discussed below.

\subsection{Transition from \ch{H2O}-dominated to CO-dominated atmospheres}

In Fig. \ref{fig:trendsvsZ} we show the evolution of the distribution of oxygen between the different O-bearing molecules as a function of the equilibrium temperature in the six planet formation scenarios from Paper I. As discussed previously, the fraction of O in the form of \ch{SiO} is virtually constant at about 6\% for all equilibrium temperatures and planetary metallicity values. 

O-bearing molecules with Mg and Fe carry a significant fraction of oxygen in the temperature range of warm giant planets but their role sharply decreases when $T_{eq}$ exceeds 800 K. At about 900 K, the contribution of \ch{CO} becomes comparable to the individual ones of Fe, Mg, and Si (see Fig. \ref{fig:trendsvsZ}). Above this temperature, CO rapidly increases while \ch{H2O}, Fe- and Mg- oxides decrease. Due to the non-stellar composition of the giant planets reported in Tab.~\ref{tab:elemental_enrichments}, the crossing point between \ch{CO} and \ch{H2O} changes with the planetary metallicity. Specifically, the crossing point shifts by about 200 K, going from 1200 K for solar-metallicity planets ($Z=1$) to less than 1000 K for the highest metallicity $Z=7.6$ (see Fig. \ref{fig:trendsvsZ}).

Furthermore, the relative importance of CO and \ch{H2O} (i.e. how much the two curves diverge at the highest temperatures) shows a non-monotonic evolution for increasing planetary metallicity (see Fig. \ref{fig:trendsvsZ}). The difference between the percentages of O as \ch{CO} and \ch{H2O} is almost zero for $Z=1$. This means that the O not sequestered by refractories equally distributes between water and carbon monoxide. Said difference sharply increases moving to $Z=1.3$, where the water accounts for less than 40\% of O and carbon monoxide about 60\% (see Fig. \ref{fig:trendsvsZ}).

Moving toward increasing values of the planetary metallicity, the imbalance in the distribution of O among CO and \ch{H2O} decreases until Z=4.8 and increases again at Z=7.6 where water accounts for about 40\% of O while carbon monoxide  accounts for about 50\%. Such non-monotonic behaviour, as well as the non-monotonic trend of water between 900 and 1000 K in Fig. \ref{fig:trendsvsZ}, arise from the different growth of the abundance of C, O, and refractories with planetary metallicity shown in Tab.~\ref{tab:elemental_enrichments}.

\input{trendsvsZ}

%% file: exoplanetdef.tex
\begin{figure*}
    \centering
    \includegraphics[]{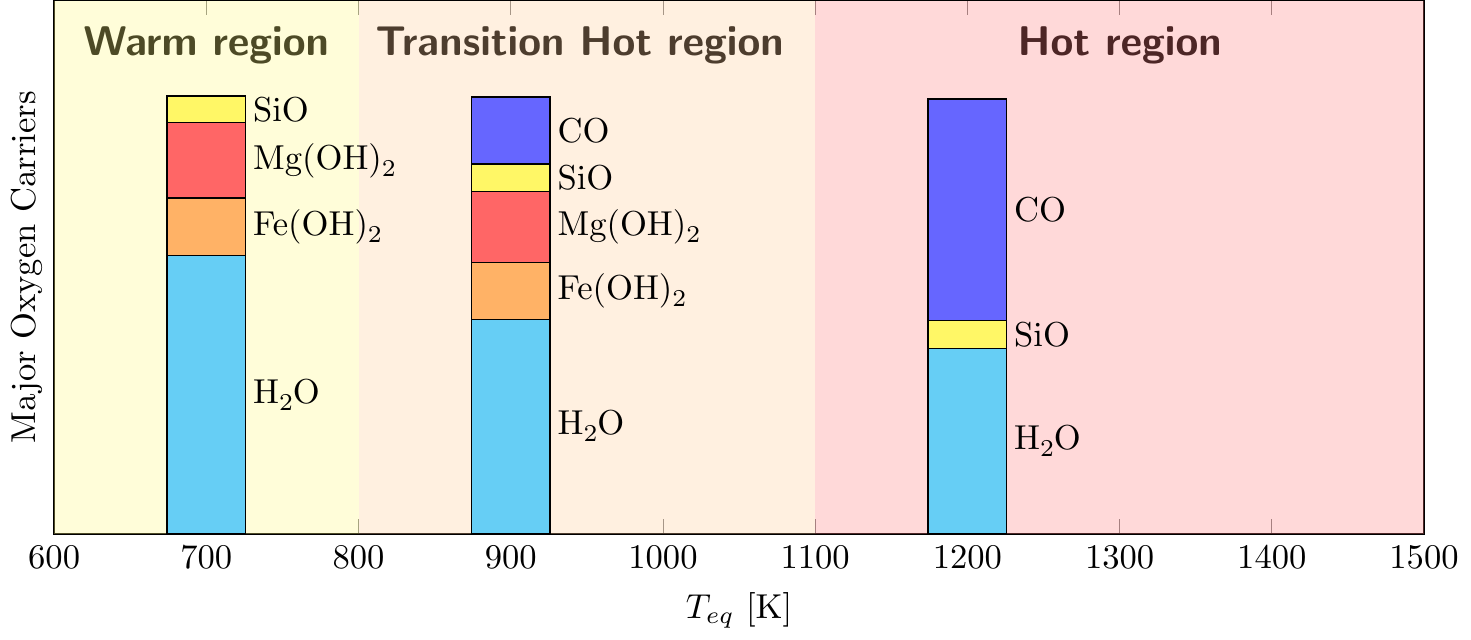}
    \caption{Illustrative example of the evolution of the main oxygen carriers going from warm to transition hot and hot giant planets. The transition hot region marks the shift between the warm temperature regime where water and refractories are the main oxygen carriers and the hot temperature regime where O is mainly in the form of \ch{CO} and \ch{H2O}.}
    \label{fig:exoplanetdef}
\end{figure*}

%% file: oxygendist-700K-800K.tex
\begin{figure*}
    \centering
    \subfigure[Exoplanet pressure-temperature profile with $T_{eq}$   @ 700 K]{
        \includegraphics[]{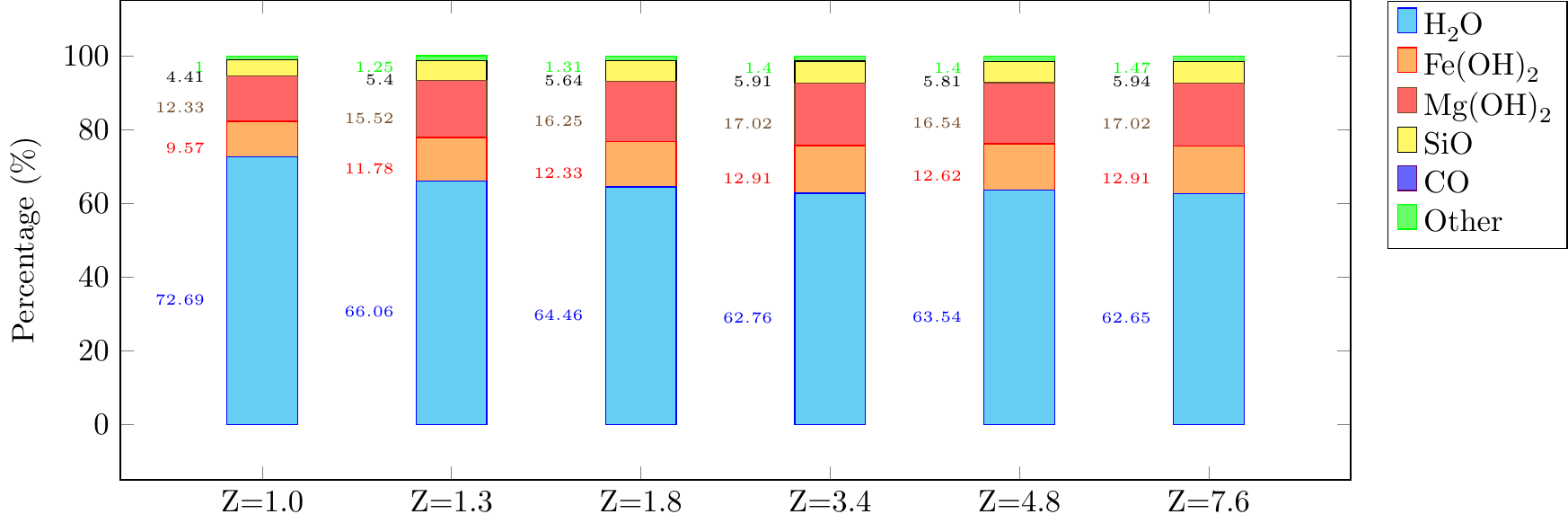}}
    \subfigure[Exoplanet pressure-temperature profile with $T_{eq}$   @ 800 K]{
        \includegraphics[]{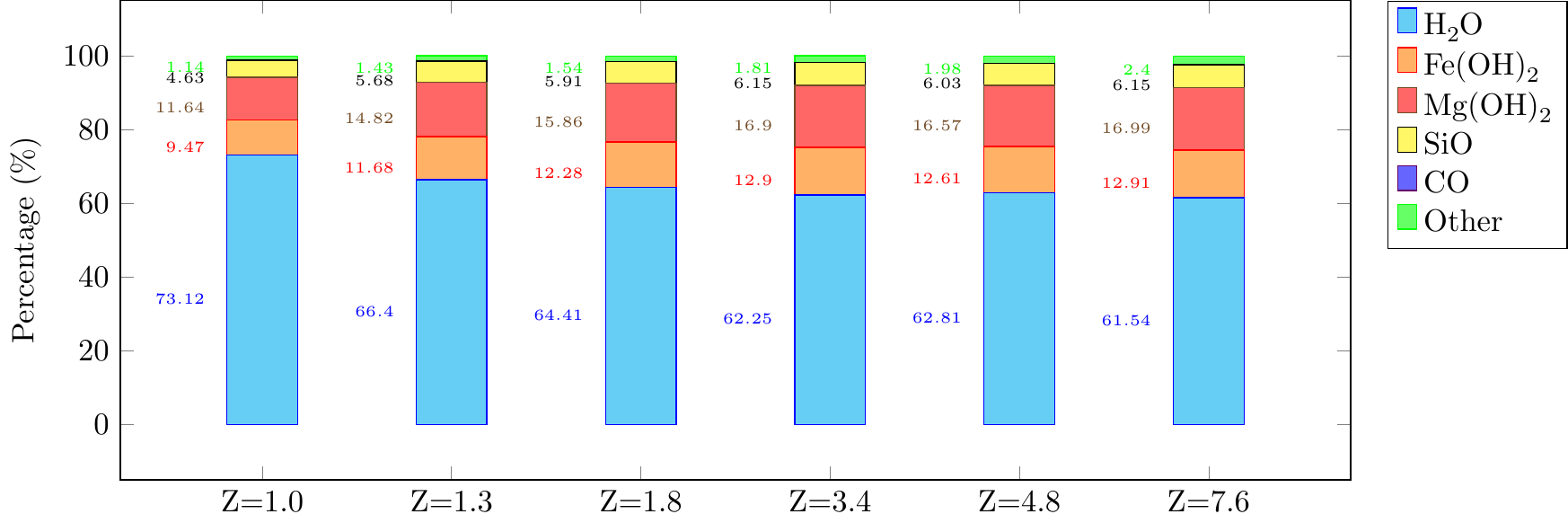}}
    \caption{Distribution of oxygen-bearing molecules in the pressure range where JWST and Ariel  have the highest sensitivity ([0.01, 1] bar) for $T_{eq}$=[700, 800] K in panels $a$ and $b$, respectively. We explicitly report only the molecules carrying a fraction of O greater than 1\%.}
            \label{fig:oxygendist1}
\end{figure*}

%% file: oxygendist-900K-1100K.tex
\begin{figure*}
    \centering
    \subfigure[Exoplanet pressure-temperature profile with $T_{eq}$   @ 900 K]{
        \includegraphics[]{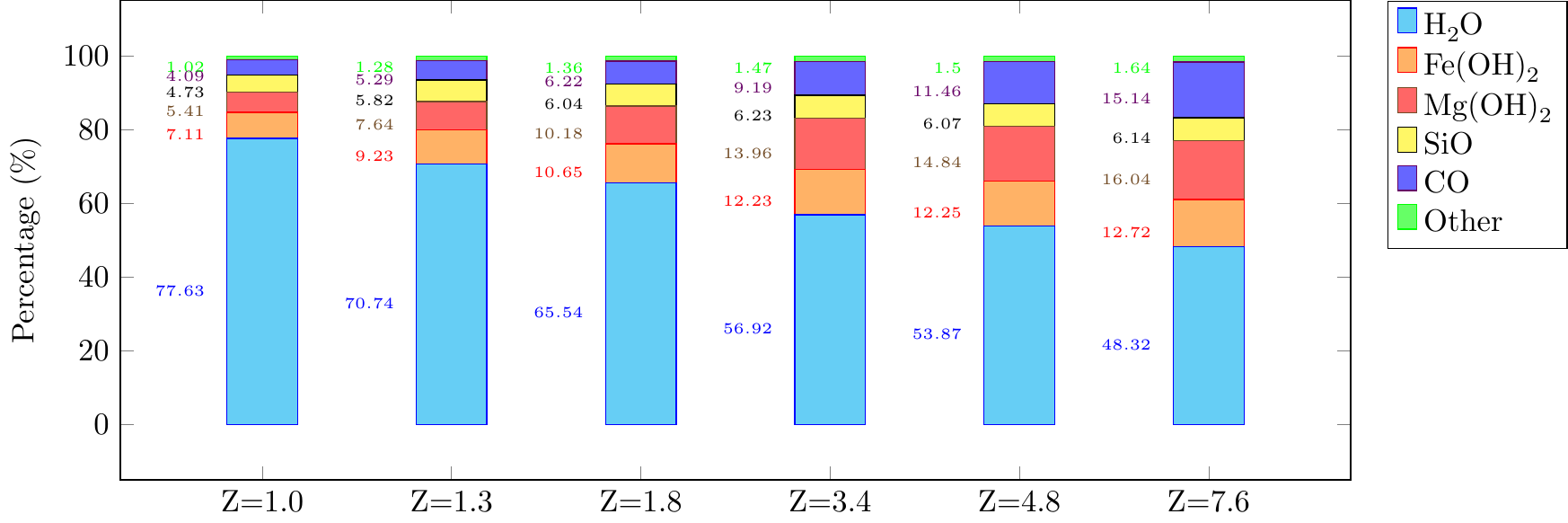}}
    \subfigure[Exoplanet pressure-temperature profile with $T_{eq}$   @ 1000 K]{
        \includegraphics[]{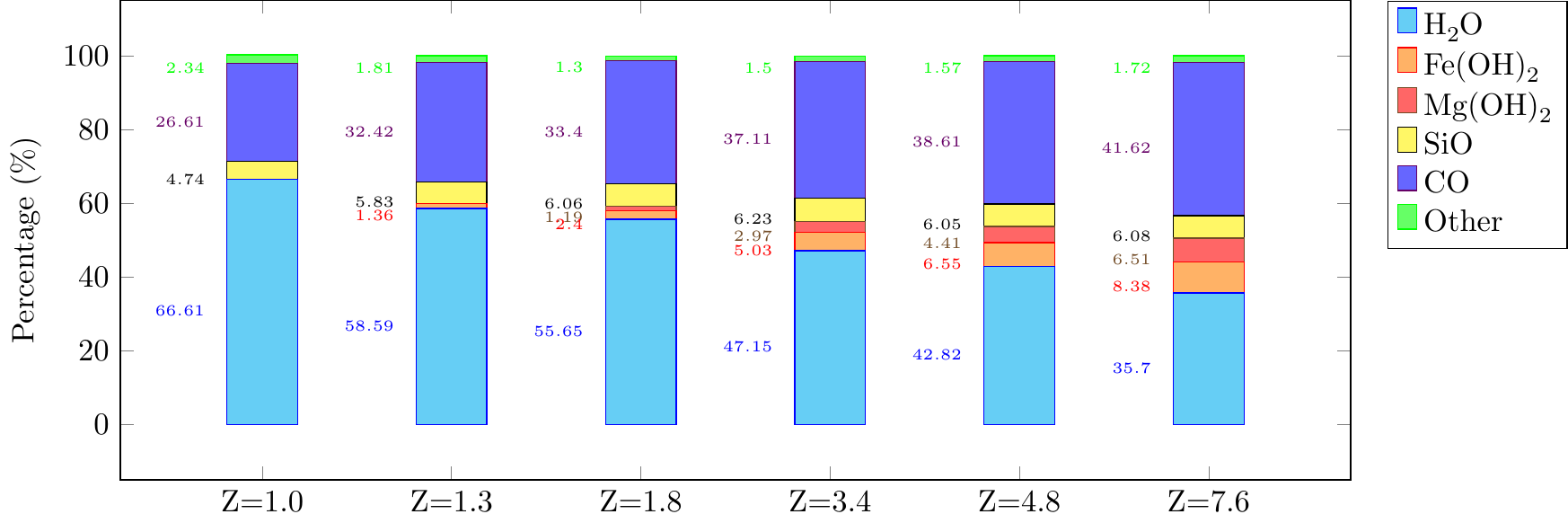}}
    \subfigure[Exoplanet pressure-temperature profile with $T_{eq}$   @ 1100 K]{
        \includegraphics[]{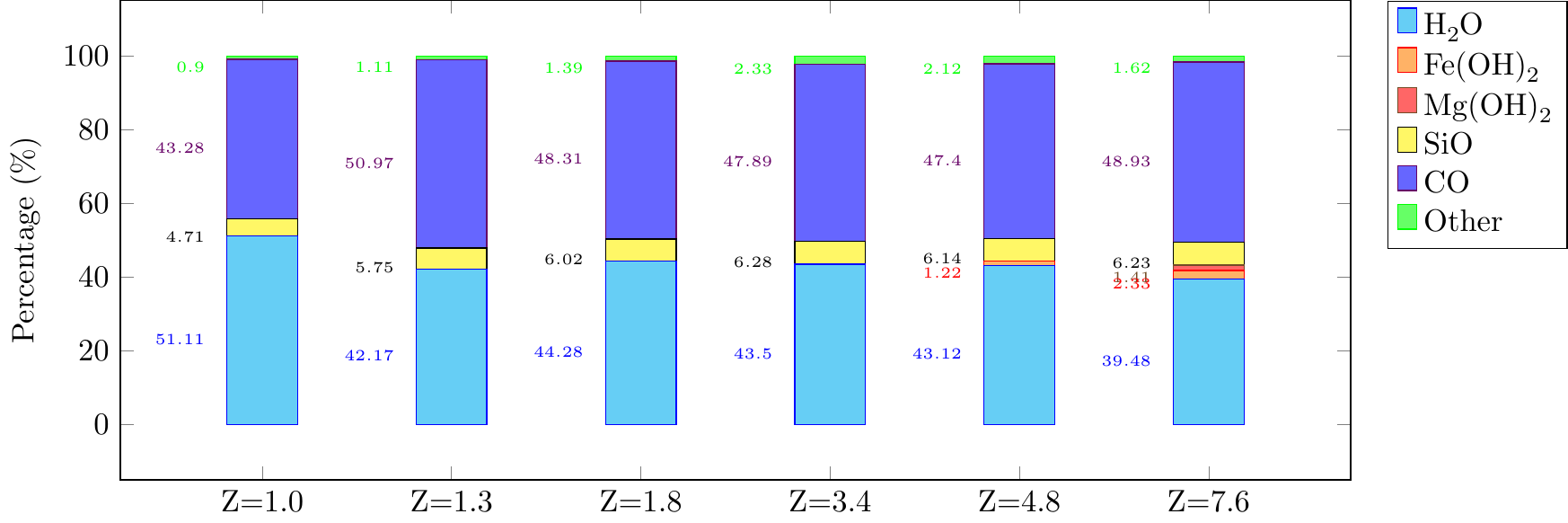}}
    \caption{Distribution of oxygen-bearing molecules in the pressure range where JWST and Ariel  have the highest sensitivity ([0.01, 1] bar) for $T_{eq}$=[900, 1000, 1100] K in panels $a$, $b$ and $c$, respectively.  We explicitly report only the molecules carrying a fraction of O greater than 1\%.}
            \label{fig:oxygendist2}
\end{figure*}

%% file: oxygendist-1200K-1500K.tex
\begin{figure*}
    \centering
    \subfigure[Exoplanet pressure-temperature profile with $T_{eq}$   @ 1200 K]{
        \includegraphics[]{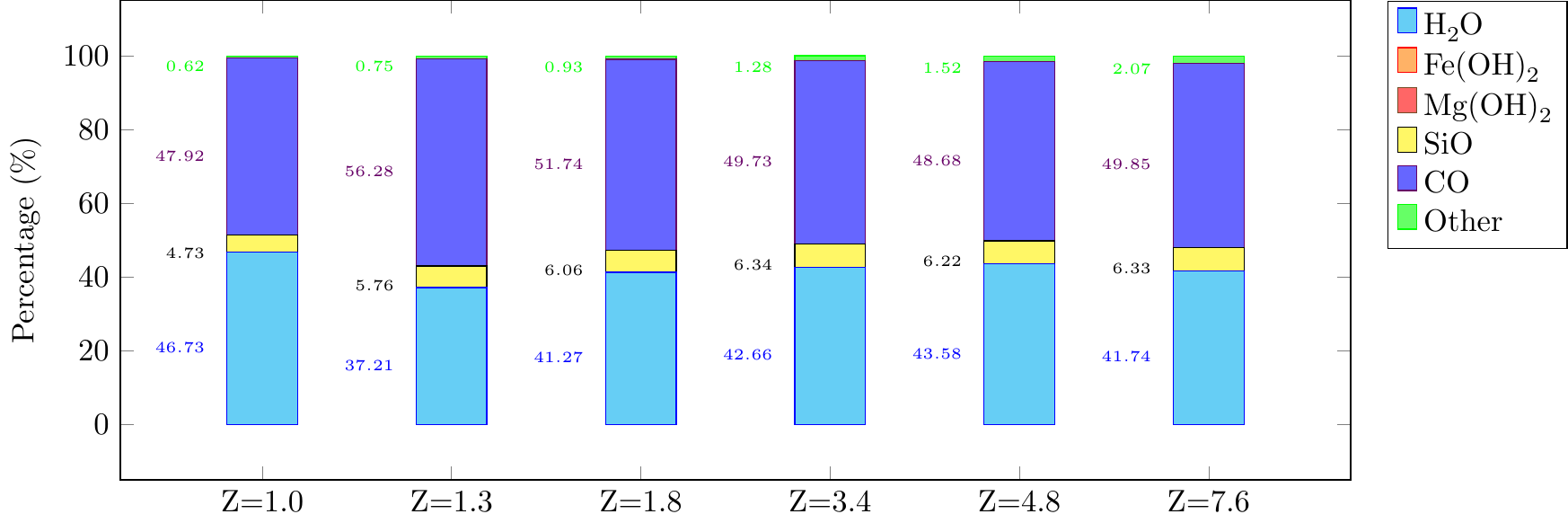}}
    \subfigure[Exoplanet pressure-temperature profile with $T_{eq}$   @ 1300 K]{
        \includegraphics[]{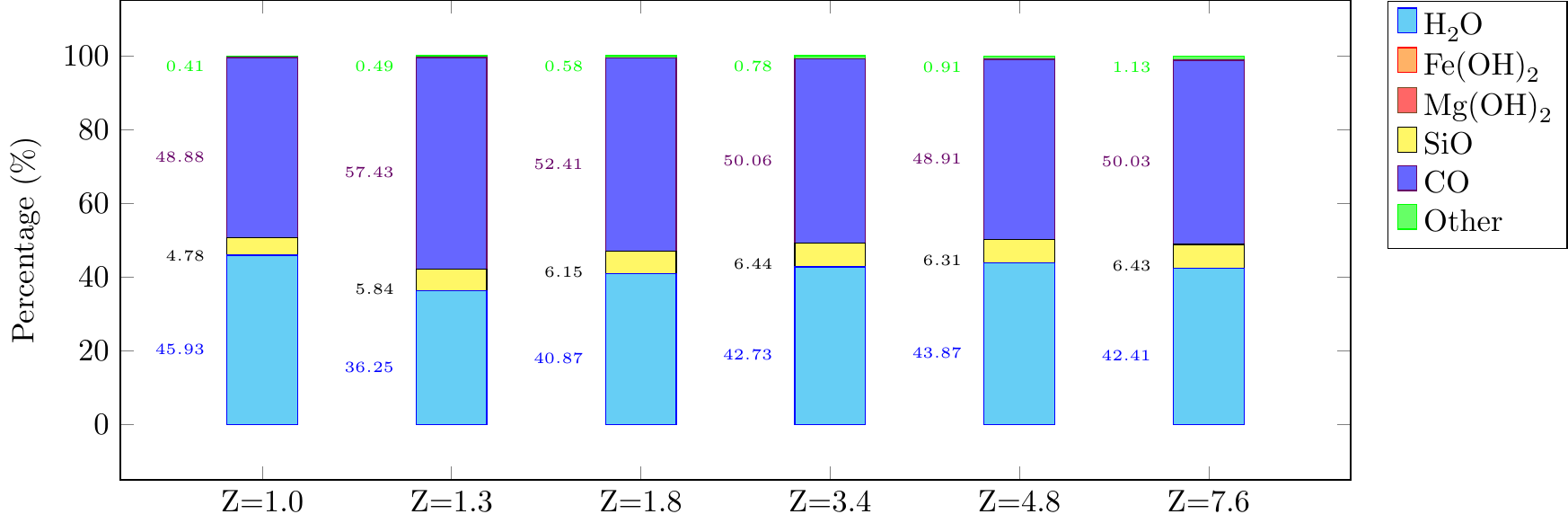}}
    \subfigure[Exoplanet pressure-temperature profile with $T_{eq}$   @ 1500 K]{
        \includegraphics[]{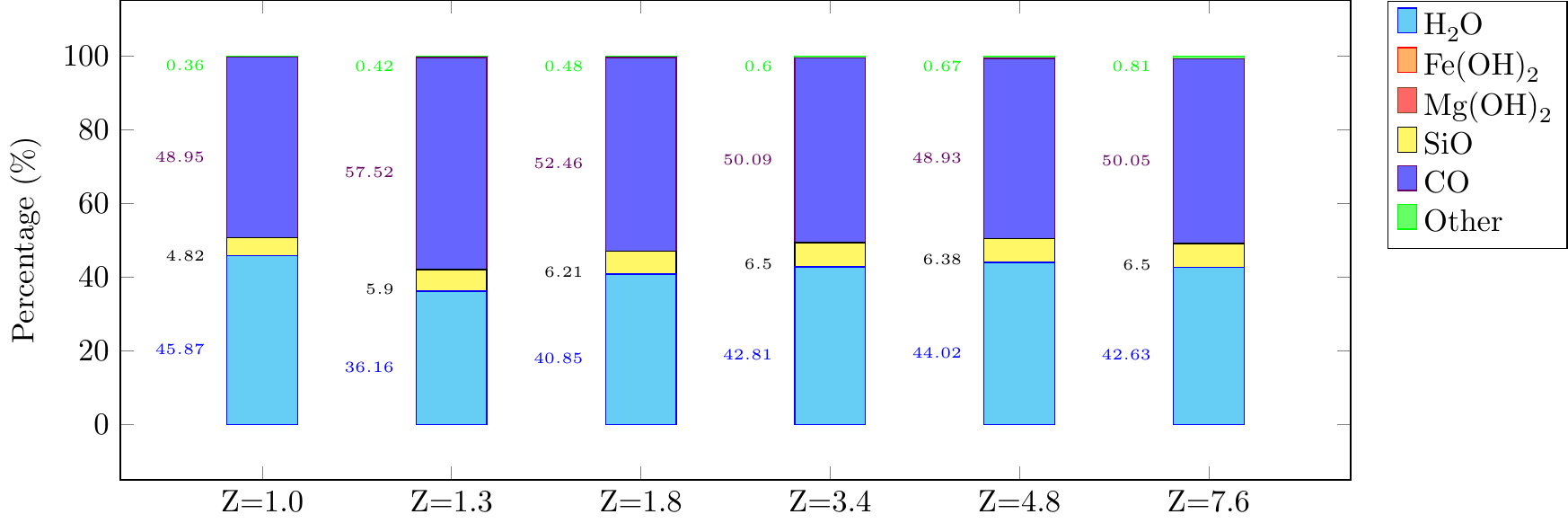}}
    \caption{Distribution of oxygen-bearing molecules in the pressure range where JWST and Ariel  have the highest sensitivity ([0.01, 1] bar) for $T_{eq}$=[1200, 1300, 1500] K in panels $a$, $b$ and $c$, respectively.  We explicitly report only the molecules carrying a fraction of O greater than 1\%.}
            \label{fig:oxygendist3}
\end{figure*}

%% file: trendsvsZ.tex
\begin{figure*}
    \centering
    \includegraphics[]{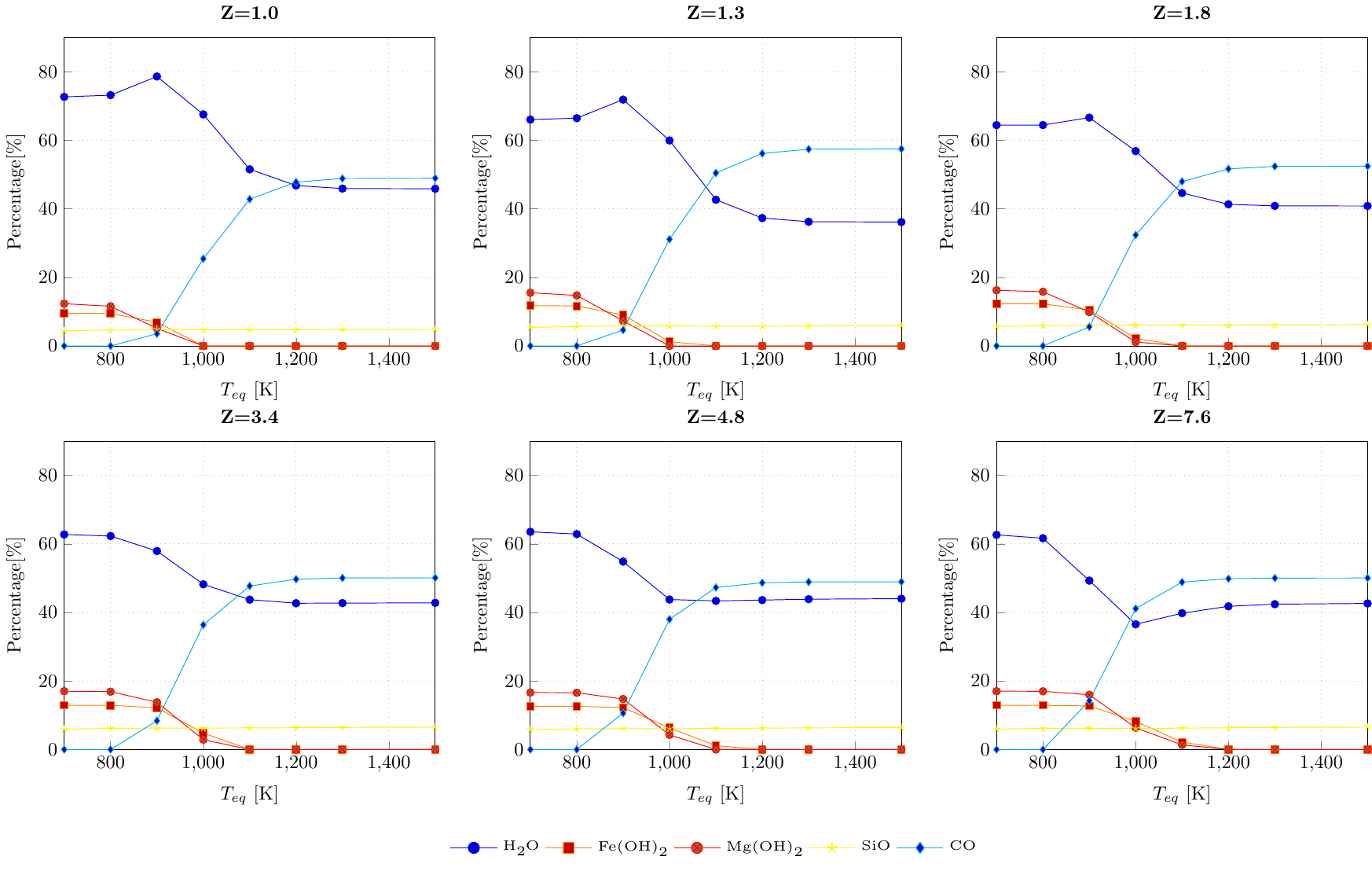}
    \caption{Evolution of the relative contributions of the five major O-bearing molecules as a function of the equilibrium temperature in the six formation scenarios. The highlighted regions mark the crossing of the \ch{CO} and \ch{H2O} curves, i.e. the temperature interval where \ch{CO} becomes the major O carrier. This crossing point shifts towards lower equilibrium temperatures as the metallicity increase.}
    \label{fig:trendsvsZ}
\end{figure*}

%% file: discussion.tex
\section{Discussion} \label{sec:discussion}
The results described in Sect. \ref{sec:results} highlight how the presence of refractory elements among the carriers of O causes the atmospheres of warm and transitional hot giant planets to be less rich in water than expected. For these planets, therefore, the role of refractories needs to be accounted for to produce accurate estimates of the atmospheric O/H abundance and C/O ratio. 

We illustrate the effect of neglecting refractory oxides on the C/O ratio in Fig. \ref{fig:oxygendeficit}, showing the trend of the oxygen deficit as a function of the metallicity and the equilibrium temperature of the giant planet. We quantify the oxygen deficit using the following formula:
\begin{equation}\label{eq:deficit}
    d_O = 1 - \frac{r}{r_s}
\end{equation}
where $r$ is the C/O ratio calculated considering all O-bearing species present in exoplanetary atmospheres. The parameter $r_s$ is the C/O ratio estimated when the only carriers for oxygen taken into account are \ch{H2O} and \ch{CO} so that
\begin{equation}
    r_s = \frac{\ch{CO}+\ch{CH4}}{\ch{H2O}+\ch{CO}}
\end{equation}    

As summarised in Fig. \ref{fig:oxygendeficit}, at fixed planetary metallicity Z the oxygen deficit $d_O$ is inversely correlated to the equilibrium temperature $T_{eq}$. In parallel, at fixed $T_{eq}$ the oxygen deficit is directly correlated to the planetary metallicity. 
At temperatures higher than 1200 K, the oxygen deficit can be approximated as constant at more or less about 6\%. Between 1000 and 1100 K, $d_O$ grows almost linearly by a factor of 3  going from about 7\% to 21\% when moving from Z=1 to Z=7.6. For decreasing equilibrium temperatures below 1000 K, the oxygen deficit can easily span between about 20\% and 40\% due to the large contribution of refractory species discussed in Sect. \ref{sec:results}.

The importance of properly accounting for the oxygen deficit is immediately highlighted by the following example. Giant planets whose metallicity is shaped by the accretion of planetesimals in the simulations of Papers I and II have Z$>$1 and C/O$\approx$0.55. An oxygen deficit d$_{O}$=0.3 (33\%, well within the range of values shown in Fig. \ref{fig:oxygendeficit}) would cause the same giant planets to appear as possessing C/O=0.8. This C/O value, however, is compatible with giant planets whose metallicity originates from the accretion of gas instead of planetesimals, meaning that not accounting for the oxygen deficit leads to incorrect constraints on the formation history.  

The correlation between oxygen deficit and planetary metallicity, however, is not constant for changing temperatures. Colder and lower metallicity planets can be characterised by the same oxygen deficit as warmer but higher metallicity planets. As a result, an uncertainty of 100 K in the planetary temperature can easily translate into an inaccuracy $\geq50\%$ in the oxygen deficit for giant planets characterised by equilibrium temperatures around 1000 K. This, in turn, can critically impact the evaluation of the C/O ratio and the assessment of the giant planet formation history.

\input{oxygendeficit}

\subsection{Implications for Jupiter in the Solar System }

The results discussed in this work impact also the study of Jupiter in the Solar System, whose atmosphere has been compositionally characterised by the NASA missions Galileo and Juno \citep{Atreya_2018,Li_2020,Grassi_2020}. Specifically, the in-situ measurements by the mass spectrometer onboard Galileo's atmospheric probe show that Jupiter's C and S are 4 and 3 times more enriched than the Sun, with 1-$\sigma$ uncertainties of about 20\% \citep{Atreya_2018}.

Jupiter's atmospheric water abundance has been recently estimated by the microwave radiometer onboard the Juno mission \citep{Li_2020}, revealing that the O abundance associated with \ch{H2O} is 2.7 times the solar one. While the measurement of the O abundance is still affected by large uncertainties (the 1-$\sigma$ uncertainty is least 60\%, \citealt{Li_2020}), this estimate suggests an atmospheric C/O ratio of 0.8. This, in turn, would point to Jupiter's heavy elements having been accreted through the disc gas \citep{bosman2019,schneider2021}. 

The observed enrichment in S, however, points to a large abundance of refractory elements and a significant role of oxygen sequestration by refractories. Since S can be used as a proxy for the enrichment of refractory species, Jupiter's atmospheric composition is similar to the scenario with Z=3.4 from Paper I\footnote{This scenario is characterised by a lower N abundance than Jupiter's nominal values from \cite{Atreya_2018} and \cite{Li_2020}, but this has no impact on the O deficit.}. To more accurately assess the oxygen deficit that should be expected for Jupiter, we reprocessed all scenarios with FastChem using the same pressure-temperature profile as \cite{Grassi_2020} based on the Galileo Entry Probe measurements \citep{Seiff_1998}.

This pressure-temperature profile is shown in Fig. \ref{fig:oxygendistjupiter} and is associated with $T_{eq}\sim 122$ K. The temperature of the atmospheric layer probed by Juno's instruments is about 260 K \citep{Seiff_1998}, as highlighted in the left-hand panel of Fig. \ref{fig:oxygendistjupiter}. Reprocessing the six planetary compositions from Sect. \ref{sec:model} with Jupiter's pressure-temperature profile results in the oxygen deficits reported in the right-hand panel of Fig. \ref{fig:oxygendistjupiter}, where we highlight the scenario with the metallicity more closely matching Jupiter's atmospheric value \citep{Atreya_2018}.

The atmospheric mixture with Z=3.4 is associated with an oxygen deficit of 32\%, meaning that water only accounts for 68\% of total oxygen. Once we correct for the oxygen deficit, Jupiter's oxygen abundance with respect to H becomes 4 times that of the Sun. This, in turn, means that Jupiter's C/O ratio becomes equal to the solar value of 0.55. This value points to Jupiter's heavy elements mainly originating from the accretion of planetesimals \citep{Turrini_2021,pacetti2022} and thus argues for a radically different formation history.

\input{oxygendist-Jupiter}

%% file: oxygendeficit.tex
\begin{figure}
    \centering
        \includegraphics[]{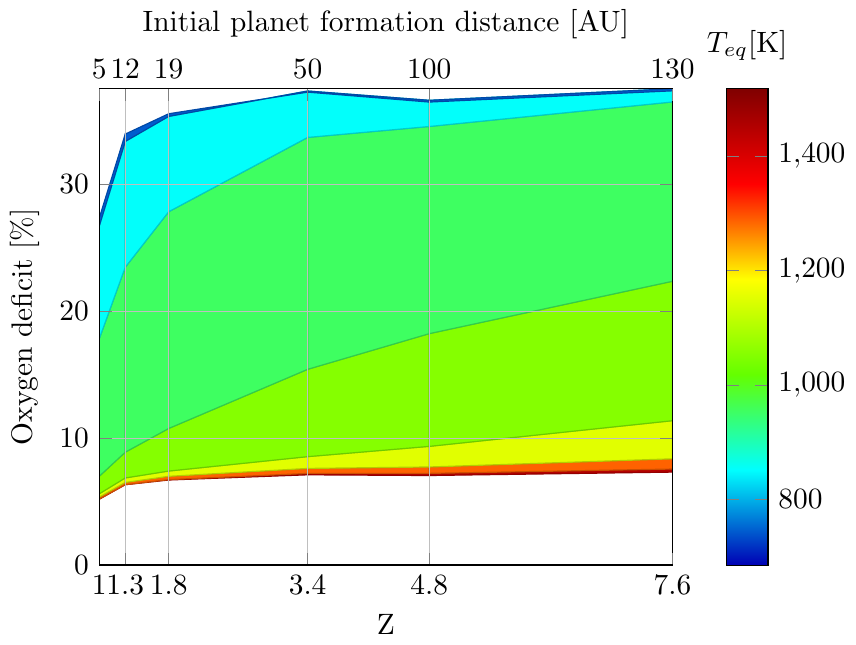}
        \caption{Oxygen deficit as a function of the planetary metallicity $Z$ and the equilibrium temperature $T_{eq}$ of the exoplanet. The oxygen deficit is defined by Eq. \ref{eq:deficit} and quantifies the systematic error introduced by accounting only for \ch{CO} and \ch{H2O} as O carriers in the planetary atmosphere.}
        \label{fig:oxygendeficit}
\end{figure}

%% file: oxygendist-Jupiter.tex
\begin{figure*}
    \centering   
        \subfigure[Jupiter pressure temperature profile]{
            \includegraphics[]{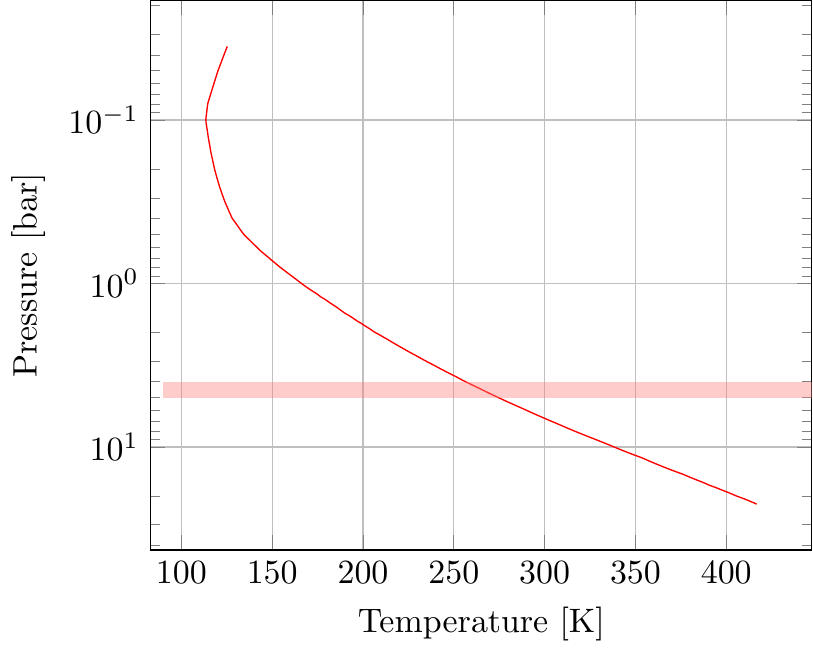}}
        \hfill
        \subfigure[Deficit trend with the metallicity]{
            \includegraphics[]{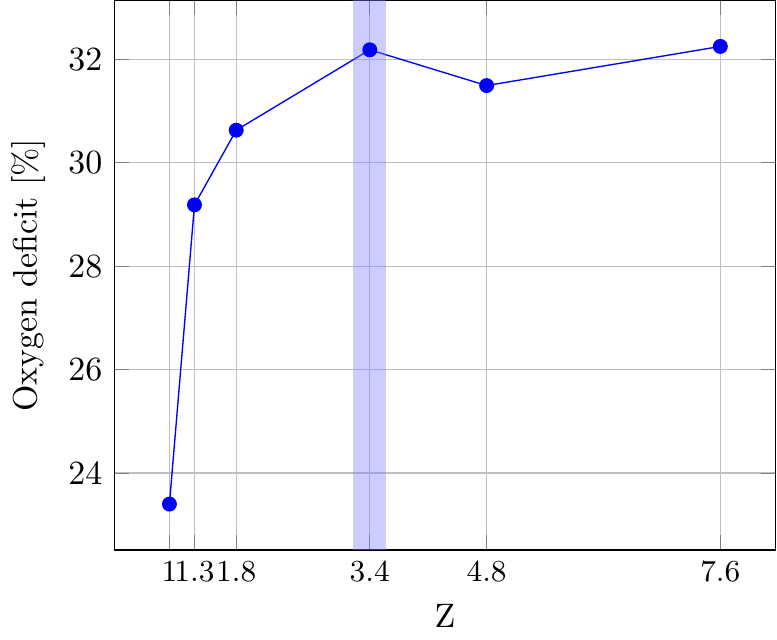}}    
        \caption{Constraints on the oxygen deficit of Jupiter. \textit{Left}: P-T profile of the Jovian atmosphere adopted from \citealt{Grassi_2020}. The highlighted red region marks the atmospheric layer probed by Juno's instruments \citep{Li_2020,Grassi_2020}. \textit{Right}: oxygen deficit of the six formation scenarios we consider in this work for the Jovian P-T profile. The highlighted blue region marks the scenario with the metallicity value closer to Jupiter's one \citep{Atreya_2018}.}
        \label{fig:oxygendistjupiter}
\end{figure*}

%% file: conclusions.tex
\section{conclusions} \label{sec:conclutions}
In this work, we explore the role of refractory elements in sequestering oxygen in the atmospheres of giant planets and its impact on estimating the atmospheric C/O ratio. We model the atmospheric chemistry assuming chemical equilibrium and using realistic elemental mixtures produced from planet formation simulations as input. These elemental mixtures are the result of the interplay between the concurrent accretion of planetesimals and disc gas by the growing giant planets and are characterised by non-solar abundance ratios between C, O, and refractory elements.

We find that the oxygen deficit depends on both the atmospheric metallicity and equilibrium temperature and, in general, does not match the classical value of 22\% estimated assuming solar elemental ratios \citep{burrows1999,Fegley_2010}. At equilibrium temperatures lower than 1000 K, the oxygen deficit can reach values of 30-40\% in the case of giant planets with high metallicity. At higher temperatures, the oxygen deficit is limited to 5-10\%, mainly due to the contribution of silicon oxides.

We also find that the interplay between atmospheric metallicity and equilibrium temperature introduces degeneracies in the oxygen deficit at temperatures close to 1000 K. Specifically, colder and lower metallicity giant planets can be characterised by the same oxygen deficit as hotter but higher metallicity planets. As shown by Fig. \ref{fig:oxygendeficit}, a 10\% uncertainty on the atmospheric temperature (i.e. about 100~K at 1000~K) introduces uncertainties of more than a factor of three in the oxygen deficit. This issue can be mitigated by observationally constraining the atmospheric abundances of oxygen and refractories or the refractory-to-oxygen ratio. Future studies will need to assess the impact of the condensation of refractory materials and cloud formation on constraining the oxygen deficit, particularly for warm and transition hot giant planets.

Our results highlight how not accounting for the oxygen deficit introduces systematic biases in quantifying the atmospheric C/O ratio of giant planets rich in refractory elements. These biases could be less marked for giant planets that accrete disc gas highly enriched in C and O by the pebble sublimation process \citep{bosman2019,schneider2021} or could be higher than estimated in this work if the giant planets accrete large amounts of oxygen-depleted planetesimals (e.g. closer to the host star). Similarly, different astrochemical environments of circumstellar discs \citep{eistrup2016,pacetti2022} could impact the magnitude of the oxygen deficit. Future studies will need to explore the role of oxygen deficit across a larger parameter space to shed light on these effects. 

Independently on these uncertainties, the results of this work highlight how ignoring the effects of oxygen deficit can lead to misinterpreting the formation history of the observed giant planets. As an illustrative example, an oxygen deficit of 30\% makes a giant planet with C/O=0.5 appear like it possesses C/O=0.8. These two values point to radically different accretion histories and sources of metallicity: the accretion of planetesimals for the first one, the accretion of disc gas 
for the second one (see Paper I and II and \citealt{schneider2021} for discussion). Adopting the second, incorrect, value as the true one, therefore, provides wrong constraints on the formation history and the native environment of the giant planet.

Finally, we apply the same methodology used for giant exoplanets to the case of Jupiter in the Solar System, taking advantage of the constraints on its abundance of oxygen and refractories and its pressure-temperature profile provided by the NASA missions Galileo and Juno. The measured atmospheric enrichment of \ch{H2O} suggests that Jupiter's oxygen abundance is 2.7 times the solar one. However, the observed abundance of sulphur, which we use as a proxy for the refractory elements, points to oxygen deficit values of the order of 30\%. After correcting for this deficit, Jupiter's oxygen abundance increases to 4 times the solar one, i.e. the same enrichment observed for carbon. This brings Jupiter's C/O ratio to match the solar value and points to the accretion of planetesimals as the source of Jupiter's heavy elements \citep{Turrini_2021,pacetti2022}.

%% file: acknowledgements.tex
\section*{Acknowledgements}
The authors acknowledge the support of the European Research Council via the Horizon 2020 Framework Programme ERC Synergy ``ECOGAL'' Project GA-855130, of the Italian National Institute of Astrophysics (INAF) through the INAF Main Stream project ``Ariel and the astrochemical link between circumstellar discs and planets'' (CUP: C54I19000700005), and of the Italian Space Agency (ASI) through the ASI-INAF contracts No. 2016-23-H.0 and 2021-5-HH.0. This project also received funding from the European Research Council (ERC) under the European Union’s Horizon 2020 research and innovation programme (grant agreement No 758892, ExoAI) and from the Science and Technology Facilities Council (STFC) grant ST/S002634/1 and ST/T001836/1 and from the UK Space Agency grant ST/W00254X/1. Danae Polychroni is supported by INAF through the project PRIN INAF 2019 ``Planetary systems at young ages (PLATEA)'' and by the Istituto Nazionale di Oceanografia e di Geofisica Sperimentale (OGS) and CINECA through the programme ``HPC-TRES (High Performance Computing Training and Research for Earth Sciences)'' award number 2022-05. Quentin Changeat is funded by the European Space Agency under the 2022 ESA Research Fellowship Program. Eugenio Schisano acknowledges the contribution from PRIN INAF 2019 through the project ``HOT-ATMOS''. The authors wish to thank Aldo Bonomo and Matteo Brogi for their discussion and feedback on exoplanetary atmospheric observations. The computational resources for this work were supplied by the Genesis cluster at INAF-IAPS and the technical support of Scigé John Liu is gratefully acknowledged.

%% file: dataavailabilitystatement.tex
\section*{Data Availability}
All data necessary to reproduce the atmospheric models are available in the article. The \textsc{FastChem} code used in the analysis is publicly available at \url{https://github.com/exoclime/FastChem}. The outputs of the planet formation simulations from \cite{Turrini_2021} and of the disc chemical models from \cite{pacetti2022} are available on reasonable request to the relevant corresponding authors. All information needed to reproduce the planet formation simulations is described in \cite{Turrini_2021}.

%% file: appendix.tex
\appendix

\section{Giant Planet Formation}\label{appendixa}

The giant planets simulated in Paper I begin their formation as planetary embryos of 0.1 M$_\oplus$ at different positions within their natal protoplanetary disc and end their growth and migration as 1 Jovian mass planets orbiting at 0.4 au from the host star. This close to the host star, further inward migration does not contribute to the compositional evolution of the giant planets in any significant way. The final orbital distance in the simulations was therefore chosen for reasons of computational efficiency (see Paper I for details).

\begin{figure*}
	\includegraphics[width=\textwidth]{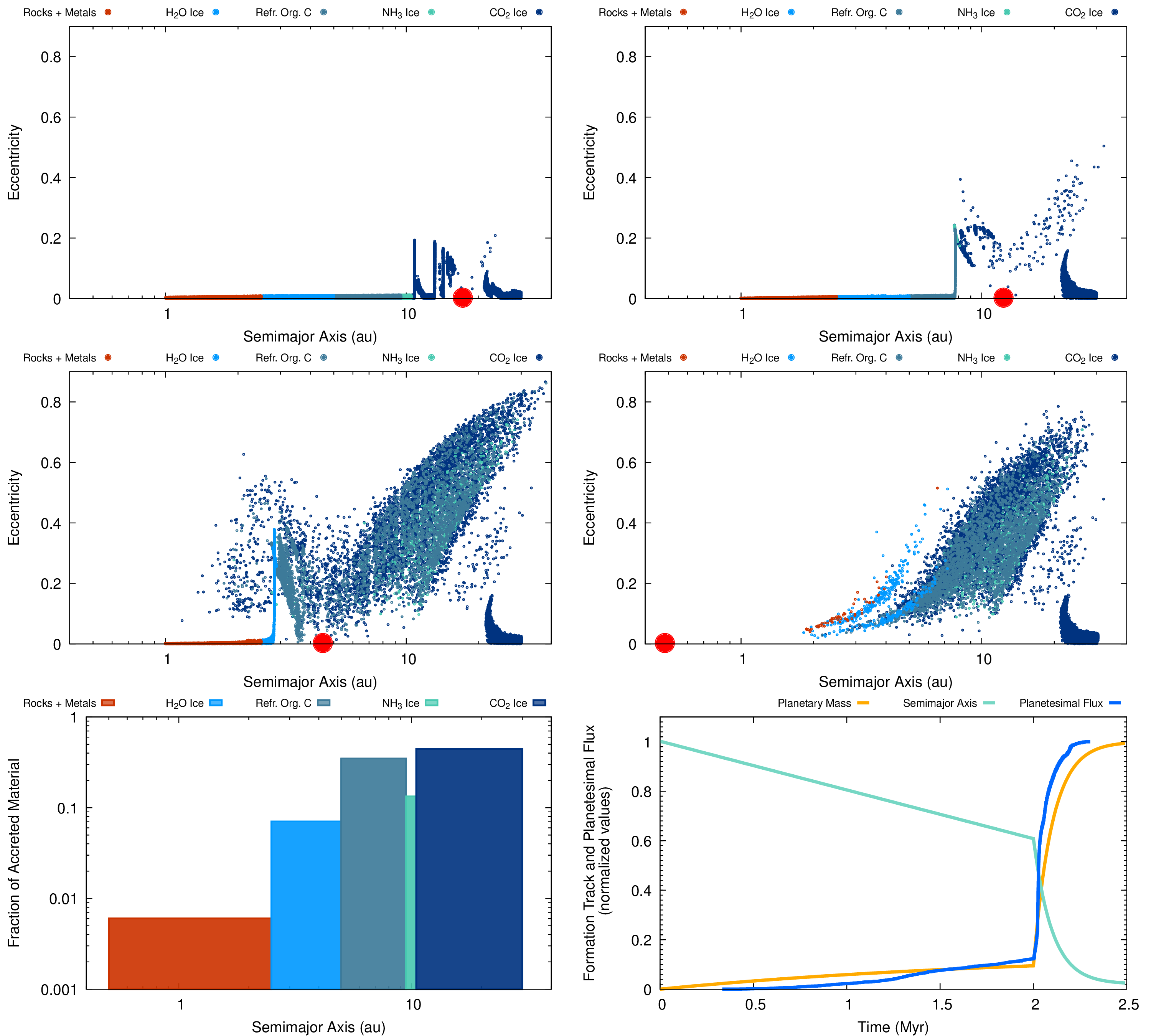}
	\caption{Formation and migration tracks of the giant planet starting its growth at 19 au in the simulations from Paper I. The first two rows show the dynamical evolution of the planetesimals in response to the growth and migration of a giant planet (large red circle) at 0.5, 1.8, 2.1, and 2.5~Myr. The different colours mark planetesimals formed in different compositional regions of the disc (the legend reports the most volatile condensate of each region). The bottom left panel shows the relative contribution of the different compositional regions in the disc to the planetesimals accreted by the giant planet. The bottom right plot shows the temporal evolution of the mass of the giant planet (orange curve), its accretion of planetesimals (blue curve), and its semimajor axis (green curve). Mass and planetesimal flux are normalised to their final values, the semimajor axis to the initial one. Figure from \citealt{pacetti2022}, who also supply an animated version of the figure.} 
	\label{fig:planetformation}
\end{figure*}

The simulations from Paper I consider six growth and migration scenarios, with the initial seed of the giant planet starting its formation track at 5, 12, 19, 50, 100 and 130 au from the host star. These starting positions imply that the six simulated giant planets cross different compositional regions of the protoplanetary disc and encounter different masses of planetesimals during their migration (see Fig. \ref{fig:planetformation}). The  simulations were performed with the parallel N-body code Mercury-Ar$\chi$es \citep{turrini2019,Turrini_2021}, which allows for accurate simulations of the aerodynamical and gravitational effects of the disc gas on the dynamical evolution of the planetesimals as well as the formation process of the giant planets. 

Mercury-Ar$\chi$es models the growth and the migration of the forming giant planets through a two-phases approach (see Fig. \ref{fig:planetformation}), based on the growth and migration tracks from \citet{bitsch2015}, \citet{dangelo2021} and \citet{mordasini2015}. The simulations also account for the temporal evolution of the radius of the giant planet based on the treatment and results of \cite{lissauer2009}. This means that the radius of the giant planet is set by its expanded atmosphere during the growth of the planetary core and undergoes a rapid contraction after the runaway gas accretion phase begins (see Fig. \ref{fig:planetformation}). The physical radius of the giant planet is used by Mercury-Ar$\chi$es to produce realistic impact fluxes of planetesimals.

The giant planets form and migrate within a protoplanetary disc whose gas surface density profile is modelled after that of HD\,163296's circumstellar disc, one of the best characterised circumstellar discs to date. The host star and the circumstellar disc in the simulations have masses of 1 and 0.053 M$_\odot$, respectively, and they are both characterised by solar composition. The solar composition is modelled based on the data from \cite{Asplund_2009} and \cite{Scott_I_2015,Scott_II_2015}. The disc temperature profile, which sets the position of the different snowlines, is modelled after that of the solar nebula \citep{hayashi1981}, i.e. $T=T_0 \left(r/\textrm{1\,au} \right)^{-\beta}$ where $\beta$=0.5 and $T_0$=280 K.

The chemical composition of the disc midplane is taken from \cite{pacetti2022}. The volatile fractions of N, C, and O are radially distributed across the disc between gas and ices based on the astrochemical simulations by \cite{eistrup2016}. In this work, we focus on the scenario of full chemical inheritance of the disc molecular composition from the pre-stellar phase and limited ionisation of the disc by the decay of short-lived radionuclides (``inheritance - low'' scenario from \citealt{eistrup2016}). The compositional model implemented by \cite{pacetti2022} further incorporates the contribution of rocks and refractory organics as carriers of O, C and N.

\begin{figure*}
	\includegraphics[width=\textwidth]{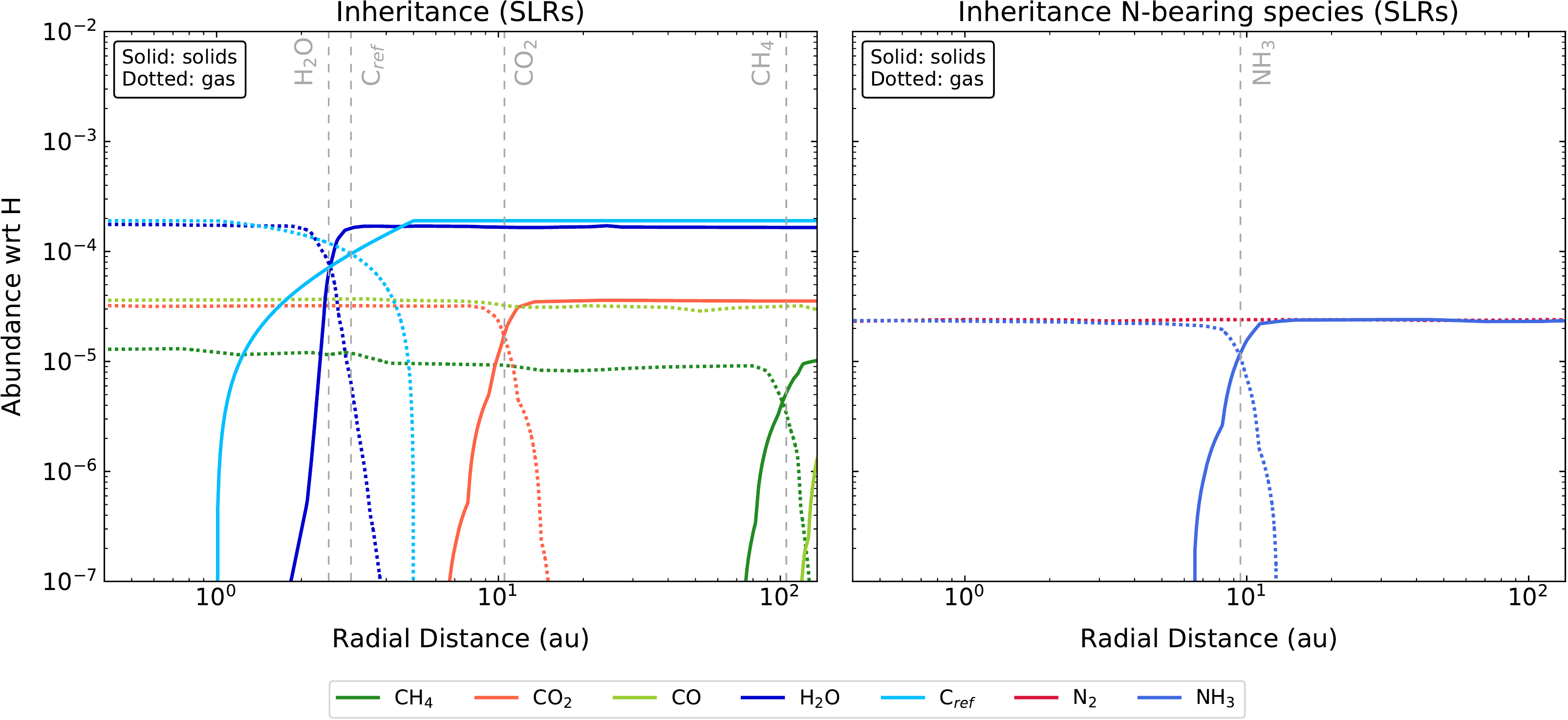}
	\caption{Disc midplane chemical structure for the volatile fractions of oxygen and carbon (\textit{left}) and of nitrogen (\textit{right}) as derived in \citealt{pacetti2022} based on the astrochemical simulations of \citealt{eistrup2016}. The left-hand plot also shows the condensation profile of refractory organic carbon according to the prescription by \citealt{cridland2019}. Based on the comparison of solar and meteoritic abundances \citep{lodders2010,Palme_2014}, 48\% of total oxygen, 9\% of carbon, 3\% of nitrogen, and the totality of S are sequestered by refractory solids (``rocks + metals'' in Fig. \ref{fig:planetformation}, see \citealt{Turrini_2021} for further discussion).}
	\label{fig:disc-chemistry}
\end{figure*}

The contribution of rocks is modelled assuming that rock-forming elements condense in the disc midplane in chondritic proportions \citep{lodders2010, Palme_2014}: the resulting mixture is identified as ``rocks + metals'' in Fig. \ref{fig:planetformation}. The term ``rock-forming elements'' encompasses all refractory elements and the fractions of O, C and N that participate in the formation of chondritic rocks. Specifically, the comparison between solar abundances and CI carbonaceous chondrites reveals that chondritic rocks carry $48\%$ of O, $9\%$ of C, and $3\%$ of N. Chondritic rocks also carry the totality of S, which we use as a proxy for all refractory elements. The major role played by refractory O revealed by meteorites is supported by the measurements of the oxygen fugacity of refractory exoplanetary material contaminating the atmospheres of polluted white dwarfs \citep{doyle2019}.

The refractory organic carbon is introduced to account for the carbon deficit observed in the Earth and solar system meteorites compared to the interstellar medium and comets (e.g. \cite{bergin2015}, and references therein). Its treatment is implemented according to the prescription used in \cite{cridland2019}, introducing a 50$\%$ condensation front at 3 au (see \citet{pacetti2022} for further details). The distribution of the volatile and refractory organic carbon across the disc, as implemented by \cite{pacetti2022} and used in this work, is shown in Fig. \ref{fig:disc-chemistry}.

We refer interested readers to \citet{Turrini_2021}, \citet{pacetti2022}, and references therein for further details on the planet formation and disc composition modelling. The distribution of elements between the different phases in the midplane sets the composition of the gas and the planetesimals accreted by the giant planets during their growth and migration. The accreted materials are reverted to their composing elements by the high temperatures of the newly formed planets \citep{lissauer2009,dangelo2021} and recombine into molecules in their atmospheres.